\documentclass[prd, twocolumn, nofootinbib, floatfix]{revtex4}

\usepackage{amsmath}
\usepackage{amsbsy}

\input epsf
\usepackage{amsmath,amssymb,subfigure}
\usepackage{graphicx}
\usepackage{epsfig}
\usepackage{color}
\usepackage{ulem}
\usepackage{epstopdf}

\newcommand{\be}{\begin{equation}}
\newcommand{\ee}{\end{equation}}

\newcommand{\beq}{\begin{equation}}
\newcommand{\eeq}{\end{equation}}
\newcommand{\beqa}{\begin{eqnarray}}
\newcommand{\eeqa}{\end{eqnarray}}

\newcommand{\ls}{\mathrel{\raise0.27ex\hbox{$<$}\kern-0.70em \lower0.71ex\hbox{{
$\scriptstyle \sim$}}}}

\begin{document} 

\title{Testing General Relativity with Current Cosmological Data} 
\author{Scott F.\ Daniel$^1$, Eric V.\ Linder$^{1,2,3}$, Tristan L.\ Smith$^3$, 
Robert R.\ Caldwell$^4$, Asantha Cooray$^5$, 
Alexie Leauthaud$^{2,3}$, Lucas Lombriser$^6$ 
} 
\affiliation{$^1$Institute for the Early Universe, Ewha Womans University, 
Seoul, Korea\\ 
$^2$Lawrence Berkeley National Laboratory, Berkeley, CA, USA\\ 
$^3$Berkeley Center for Cosmological Physics, University of California, 
Berkeley, CA, USA\\ 
$^4$Department of Physics and Astronomy, Dartmouth College, Hanover, NH, USA\\ 
$^5$Department of Physics and Astronomy, University of California, Irvine, 
CA, USA\\ 
$^6$Institute for Theoretical Physics, University of Z{\"u}rich, 
Switzerland} 

\date{\today}

\begin{abstract} 
Deviations from general relativity, such as could be responsible for the 
cosmic acceleration, would 
influence the growth of large scale structure and the deflection of light 
by that structure.  We clarify the relations between several different 
model independent approaches to deviations from general relativity appearing 
in the literature, devising a translation table. We examine current 
constraints on 
such deviations, using weak gravitational lensing data of the CFHTLS and 
COSMOS surveys, cosmic microwave background radiation data of WMAP5, and 
supernova distance data of Union2.  A Markov Chain Monte Carlo likelihood 
analysis of the parameters over various redshift ranges yields consistency 
with general relativity at the 95\% confidence level. 
\end{abstract} 

\maketitle

\section{Introduction \label{sec:intro}}

The nature of gravitation across cosmological ages and distances remains a 
frontier of current knowledge as we try to understand the origin of the cosmic 
acceleration \cite{Frieman:2008sn,Caldwell:2009ix}. Newly refined observations 
of cosmic structure \cite{Massey:2007gh,Fu:2007qq} make it possible to test the
predictions of general relativity (GR) for its influence on the growth of 
cosmic structure through gravitational instability and the gravitational 
lensing deflection of light by that structure. Indications of a deviation from 
GR would have profound consequences for cosmology, as well as for fundamental 
physics.

To explore for new gravitational phenomena, it is useful to parameterize 
the deviations from GR in the gravitational field equations.  A common 
approach is to introduce two new parameters.  The first parameter imposes 
a relation between the two gravitational potentials entering Newton's 
gravitational law of acceleration and the Poisson equation. 
These are equal in GR in the absence of anisotropic stress but different 
in many theories of modified gravity.  
The second parameter establishes a new relation between the metric and matter 
through a modified Poisson-Newton equation, which can be viewed as turning 
Newton's gravitational constant into an effective function of time and space.  
Numerous realizations of these relations have been put forward in the 
literature 
\cite{Bertschinger:2006aw,Caldwell:2007cw,Zhang:2007nk,Amendola:2007rr,Hu:2007pj,Amin:2007wi,Jain:2007yk,Bertschinger:2008zb,Hu:2008zd,Song:2008vm,Schmidt:2008hc,Skordis:2008vt,Linder:2009kq,Dent:2009wi}.  

One motivation for our study is to attempt to relate these disparate, but 
closely related, approaches. Furthermore, many studies have focused on the 
ability of future measurements to discriminate among various models and to 
carry out parameter estimation \cite{Schmidt:2007vj,Zhao:2008bn,Pogosian:2010tj,Song:2008xd,Serra:2009kp,Zhao:2009fn,Guzik:2009cm,Kosowsky:2009nc,Beynon:2009yd,Masui:2009cj,Song:2010rm,Zhang10}, 
however there is sufficient data at present to evaluate 
preliminary tests of GR \cite{Di Porto:2007ym,Nesseris:2007pa,Dore:2007jh,Daniel:2008et,Yamamoto:2008gr,Daniel:2009kr,Giannantonio:2009gi}.  
We concentrate 
here on current constraints, which also allows us to examine a recent claim 
of a possible departure from GR \cite{Bean:2009wj}.  
 
The main points of this article are thus to 1) clarify the relation between 
different parameterizations and what the degrees of freedom are in a 
consistent system of equations of motion, 2) confront the parameters 
encoding deviations from GR with current data to test the theory of gravity, 
and 3) discuss which features of the data have the most sensitivity to 
such a test and what astrophysical systematics may most easily mimic a 
deviation. 

In Sec.~\ref{sec:eqs} we lay out the gravitational field equations in terms 
of the metric potentials and matter perturbations and compare several forms 
of parameterizations, giving a ``translation table'' between them. We 
illustrate in Sec.~\ref{sec:influ} the influence of the parameters on the 
cosmic microwave background (CMB) temperature power spectrum, the matter 
growth and power spectrum, and the weak lensing shear statistics. Using 
Markov Chain Monte Carlo (MCMC) techniques, we then constrain the deviation 
parameters with current data in Sec.~\ref{sec:obs}. We briefly discuss 
astrophysical systematics and future prospects in Sec.~\ref{sec:fut}. 

\section{Systems of Parameterizing Gravity \label{sec:eqs}} 

The most accurate observations of the effects of gravity have been made in the 
local universe, e.g.\ within the solar system and in binary neutron star 
systems \cite{Bertotti:2003rm,Shapiro:2004zz,Taylor:1994zz,Lyne:2004cj}.  
These observations can be used to distinguish between various theories of 
gravity through the parameterized post-Newtonian (PPN) formalism 
\cite{Will:1993ns,Will:2005va}. The standard PPN formalism introduces a set of 
constant parameters that take on various values in different gravity theories. 
This, however, does not give a full description of possible deviations  from General Relativity over cosmological scales. 

Recent interest in modified gravity has concentrated on those theories that 
can serve as an alternative explanation for the current period of accelerated 
cosmic expansion.  In order for modifications producing late-time acceleration 
on cosmic scales to agree with local tests of gravity they must contain length 
and/or time dependent modifications, which do not occur in the standard PPN 
formalism.  Moreover, for some theories the natural arena for the PPN 
formalism -- solar system and binary neutron star system observations -- may 
be less discriminating than cosmological tests of gravity, given that the 
modifications are on large scales. This has led to efforts to establish a 
parameterized formalism that allows for meaningful comparison between modified 
gravity theories within a cosmological framework 
\cite{Bertschinger:2006aw,Caldwell:2007cw,Zhang:2007nk,Amendola:2007rr,Hu:2007pj,Amin:2007wi,Jain:2007yk,Bertschinger:2008zb,Hu:2008zd,Song:2008vm,Schmidt:2008hc,Skordis:2008vt,Linder:2009kq,Dent:2009wi}, 
without assuming a specific model.

\subsection{Degrees of Freedom \label{sec:dof}} 

Changes in the laws of gravitation affect the relationship between the 
metric and matter variables.  Let us explore the degrees of freedom 
available to define this relation. Restricting our attention to scalar degrees 
of freedom of the gravitational field, the metric has only two physically 
relevant scalar functions, or potentials, given by the line element (in 
conformal-Newtonian 
gauge, adopting the notation of \cite{Ma:1995ey}) 
\begin{equation}
ds^2=a^2\,[-(1+2\psi)\,d\tau^2+(1-2\phi)\,d\vec{x}^2]\,, 
\end{equation} 
where $a$ is the scale factor, $\tau$ the conformal time, and $x$ the 
spatial coordinate.  In addition to the metric potentials $\phi$ and 
$\psi$, perturbations to a perfect fluid introduce four additional scalar 
functions: density perturbations $\delta \rho$, pressure perturbations 
$\delta p$,  velocity (divergence) perturbations $\theta$, and a possible 
nonzero anisotropic stress $\sigma$. 

The dynamics of any particular theory 
are then specified when six independent relations between these six 
quantities are given.  Further restricting attention to those gravity 
theories that maintain the conservation of stress 
energy, 
$\nabla^{\mu} T_{\mu \nu} = 0$, the resulting generalized continuity and 
Euler equations give two scalar equations and the gravitational field 
equations supply the remaining four \cite{Ma:1995ey}.  

Since the cosmic expansion shifted from deceleration to acceleration only 
recently, since  $z<0.5$ \cite{Turner:2001mx}, gravity theories that account 
for this transition without any physical dark energy require a significant 
departure from GR at late times.  Consequently, nonrelativistic matter is the 
dominant component of the cosmological fluid and so 
$\delta p = \delta p_m = 0$ and $\sigma = \sigma_m =0$. Hence, in these 
theories the dynamically important equations consist of two, as yet 
unspecified gravitational field equations and the two equations of 
stress-energy conservation applied to matter, which in Fourier-space are given 
by
\begin{eqnarray}
\dot{\delta}_m &=&- \theta_m + 3 \dot{\phi}, \label{SE1} \\
\dot{\theta}_m &=& - \mathcal{H} \theta_m + k^2 \psi \, .   \label{SE2}
\end{eqnarray}
In the above equations, $\delta_m\equiv\delta\rho_m/\bar\rho_m$ with 
$\bar{\rho}_m$ the homogeneous part of the matter density, 
$\mathcal{H} \equiv \dot{a}/a$, the dot denotes a derivative with respect to 
conformal time, and $k$ is the wavenumber.  
There still remains freedom in setting the two gravitational field equations 
to close the system, subject to the requirement that the theory approaches 
GR within the solar system. 

The two field equations that can close the system in the case of GR are 
\begin{eqnarray}
\nabla^2 \phi &=& 4 \pi G a^2 \bar{\rho}_m \Delta_m, \label{poisson}\\
\psi &=& \phi,
\end{eqnarray}
where 
\begin{equation}
\Delta_m \equiv \delta_m+\frac{3\mathcal{H}}{k^2}\theta_m \,. 
\end{equation}

In a wide variety of alternative theories of gravitation, additional 
scalar degrees of freedom modify the strength of Newton's constant, and 
enforce a new relationship between the potentials $\phi$ and $\psi$. 
Therefore, one choice for the modified field equations in Fourier-space is 
\begin{eqnarray}
-k^2\, \frac{A \phi + B \psi}{A+B} &=& 4 \pi G \mu(\tau,k) \bar{\rho}_m 
\Delta_m \,, \\
\phi &=& \eta(\tau,k) \psi \,,
\end{eqnarray} 
where $A$ and $B$ are constants, and $\mu$ and $\eta$ are functions of time 
and scale, which are still to be determined.  As we will see, there are 
many other choices that can be made for the exact form of parameterization. 
These choices influence the constraints and the correlations between those constraints 
that particular observations give for a particular set of post-GR parameters.
We discuss 
some of the frameworks in the next subsections.

\subsection{$\varpi\mu$CDM}
\label{sec:vmucdm}

We refer to the equations of motion used in \cite{Caldwell:2007cw,Serra:2009kp,Daniel:2008et,Daniel:2009kr} 
as $\varpi$CDM.  In $\varpi$CDM, the equations of motion for cosmic 
perturbations are determined by enforcing the relation
\begin{equation}
\psi=[1+\varpi(\tau,k)]\,\phi \label{varpidef}
\end{equation}
for the potentials arising from non-relativistic matter, where the departure 
from GR is controlled by the parameter $\varpi$. In practice, this is carried 
out by adding a source to the off-diagonal space-space Einstein equation in 
order to simulate a smooth transition from GR to modified gravity. 

Next, 
requiring that the new gravitational phenomena do not introduce a preferred 
reference frame distinguished by a momentum flow, e.g.\ a $\theta$ that would 
be attributed to a dark fluid, the time-space Einstein equation is preserved, 
whereby
\begin{equation}
\label{timespace}
-k^2 \left(\dot{\phi} + \mathcal{H} \psi\right)=-4\pi Ga^2(\bar\rho+\bar{p})\theta \,.
\end{equation}
As discussed in \cite{Daniel:2009kr}, preserving the time-space Einstein 
equation along with the modification in Eq.~(\ref{varpidef}) still results 
in a correction to the GR Poisson equation. This can be thought of as a 
consequence of the conservation of stress-energy and the related Bianchi 
identity as applied to the modified gravitational field equations.  
 
We now propose to extend the $\varpi$CDM model, to incorporate a new 
parameter $\mu$ that controls the modification to the Poisson equation, i.e. 
\begin{equation}
\label{mudef}
-k^2 \phi=\mu(\tau,k)\,4\pi Ga^2\bar{\rho}_m\Delta_m \,. 
\end{equation}
Procedurally, this equation replaces Eq.~(\ref{timespace}) for obtaining the 
evolution of the gravitational fields. We call this new parameterization 
$\varpi \mu$CDM and note that in this parameterization the time-space 
Einstein equation is generally modified, as opposed to in the $\varpi$CDM 
parameterization. Note that setting $\mu=1$ does not reproduce the original 
$\varpi$CDM model since there $\varpi$ itself modifies the Poisson equation 
as discussed above.  
This parametrization is consistent with the conservation of
large-scale curvature perturbations following an argument made in 
Ref.~\cite{Pogosian:2010tj}.  They argue that super-horizon curvature
perturbations are conserved, so long as the velocity perturbation $\theta$ is of
order $(k/\mathcal{H})^2$.  This can be seen to be true from
Eq.~(\ref{thetadef}) presented in the appendix of this work.

\subsection{PPF Linear Theory}

A parameterized post-Friedmann (PPF) framework of linear fluctuations was 
introduced by~\cite{Hu:2007pj, Hu:2008zd} to describe modified gravity models 
that yield cosmic acceleration without dark energy. It captures modifications 
of gravity on horizon, sub-horizon, and non-linear scales. 
Once the expansion 
history is fixed, the model is defined by three functions and one parameter, 
from which the dynamics are derived by conservation of energy and momentum and 
the Bianchi identities. Modifications to the relationship between the two 
metric perturbations are quantified by the metric ratio
\begin{equation}
g(a,k) \equiv \frac{\phi-\psi}{\phi+\psi}\,.
\label{eq:g_ppf}
\end{equation}
In the linearized Newtonian regime, a second function $f_G(a)$ relates 
matter to metric perturbations via 
\begin{equation}
-k^2 (\phi+\psi) = \frac{8\pi G}{1+f_G}\, a^2 \bar{\rho}_m \Delta_m\,. 
\label{eq:poisson_ppf}
\end{equation}
The corresponding quantity that defines this relationship on superhorizon 
scales is $f_{\zeta}(a)$.  The last quantity 
that needs to be defined is  $c_{\Gamma}$, which determines the transition 
scale from superhorizon to quasistatic behavior in the dynamical equations 
(see \cite{Hu:2007pj, Hu:2008zd} for details).  

The PPF parameters can be directly related to the $\varpi\mu$CDM parameters as follows: 
\begin{eqnarray}
g &=& -\frac{\varpi}{2+\varpi} \quad ;\quad \varpi=-\frac{2g}{1+g}\\
f_G &=& \frac{2}{\mu(2+\varpi)} - 1 \quad ;\quad \mu=\frac{1+g}{1+f_G}\,.
\label{translation1}
\end{eqnarray}

\subsection{Gravitational Growth Index $\gamma_G$ \label{sec:gamma}} 

Another way to close the system of equations is to specify the evolution 
of one of the perturbed fluid or metric variables.  A standard choice is 
to determine a specific evolution for $\Delta_m$ through the gravitational 
growth index $\gamma_G$ introduced to parameterize deviations from general 
relativity in growth by \cite{Linder:2005in}.  This was partly tied to the 
metric potentials in \cite{Linder:2007hg} but here we present a more 
complete relation.  

From Eq.~(23) of \cite{Linder:2007hg} we see the key quantity is the 
modification of the source term in the Poisson equation, there called $Q$. 
The second order equation for the evolution of the density perturbation 
arises from $\nabla^2\psi$, and there is also a modification $\mu$ allowed 
in the gravitational coupling as in Eq.~(\ref{mudef}).  In essence, 
$\nabla^2\psi \to -k^2(1+\varpi)\phi \to (1+\varpi)\mu\times 4\pi Ga^2 
\bar{\rho}_m\Delta_m$.  Thus $Q=(1+\varpi)\mu$.  The relationship between
$\varpi$, $\mu$ and the evolution of $\Delta_m$ is presented 
rigorously here in Eq.~(\ref{Ddd}) (also see Sec.~\ref{sec:mat}). 

The gravitational growth index in Eq.~(23) of \cite{Linder:2007hg} 
thus relates to the $\varpi\mu$CDM formalism 
through 
\beqa 
\gamma_G&=&\frac{3(1-w_\infty-[(1+\varpi)\mu-1]/[1-\Omega_m(a)])} 
{5-6w_\infty} \\ 
&\to& \frac{6}{11}\,\left(1-\frac{\varpi_0+\mu_0}{2} 
\frac{\Omega_m}{1-\Omega_m}\right) \,. \label{eq:gammawu} 
\eeqa 
Note $w_\infty$ is an effective high redshift equation of state defined in 
terms of how the matter density in units of the critical density, 
$\Omega_m(a)$, deviates from unity (specifically, 
$w_\infty=[d\ln\Omega_m(a)/d\ln a]/[3(1-\Omega_m(a)]$).  
In the last line of Eq.~(\ref{eq:gammawu}) we specialize to 
a $\Lambda$CDM expansion history, as used throughout this article, so 
$w_\infty=-1$, and to the ansatz for $\varpi$ and $\mu$ used later in 
Eqs.~(\ref{paramsacubed}).

\subsection{Relating Parameterizations}

The discussion above is by no means an exhaustive list of the 
parameterizations 
proposed in the literature to describe departures from GR. Many more exist, 
and while all of them have in common a relatively simple parameterization of 
the departure from $\phi=\psi$, they all differ in how they close
the system of equations.  Some, like $\varpi\mu$CDM, modify the Poisson 
equation directly.   Others, like $\varpi$CDM, retain one of the Einstein 
equations.  

Table~\ref{comparisontable} lists some of the most common parameterizations 
and presents a useful translation between 
their post-GR parameters and $\varpi\mu$CDM.  With the possible exception of 
the parameterization from \cite{Bean:2009wj} 
(see next paragraph and its footnote), all of the 
parameterizations presented are presumed to leave the equations of 
stress-energy conservation unmodified.

\begin{table*}[!tb]
\begin{tabular}{l l l l}
&\qquad parameter
&\qquad closing&\\
parameterization&\qquad relating $\phi$ and $\psi$&\qquad parameter
&\qquad comments\\
\hline
$\varpi$CDM \cite{Caldwell:2007cw,Daniel:2008et,Daniel:2009kr}&
\qquad$\varpi$&\qquad Retains equation (\ref{timespace})&\\
\\
Curvature \cite{Bertschinger:2008zb}&
\qquad$\gamma_{BZ}=\frac{1}{1+\varpi}$&\qquad Conserves curvature perturbations
$\zeta$&\qquad Effectively retains Eq.~(\ref{timespace}).\\
&&&\qquad See appendix in Ref.~\cite{Daniel:2008et}\\
\\
PPF \cite{Hu:2007pj, Hu:2008zd}&\qquad$g=-\frac{\varpi}{2+\varpi}$&
\qquad$f_G=\frac{2}{\mu(2+\varpi)}-1$
&\qquad Includes scale-dependent transition\\
&&&\qquad between super- and sub-horizon
regimes\\
\\
MGCAMB \cite{Zhao:2008bn,Pogosian:2010tj}&\qquad$\gamma_{MGC}=\frac{1}{1+\varpi}$
&\qquad$\mu_{MGC}=\mu(1+\varpi)$&\qquad Modifies Poisson equation with $\psi$\\
cf.\ \cite{Zhang:2007nk,Amendola:2007rr,Amin:2007wi}&\qquad$\eta=\frac{1}{1+\varpi}$&\qquad$\tilde G_{\rm eff}=
\frac{\mu(2+\varpi)}{2}$&\qquad instead of $\phi$ in Eq.~(\ref{mudef})\\
\\ 
Sub-horizon \cite{Amin:2007wi}&\qquad$\frac{B}{\Gamma_\phi}=1+\varpi$&\qquad 
$\Gamma_\phi=\mu$&\\ 
\\ 
Growth index \cite{Linder:2005in}&\qquad additional &\qquad $\gamma_G=\frac{6}{11}
\left(1-\frac{\varpi_0+\mu_0}{2} \frac{\Omega_m}{1-\Omega_m}\right)$& 
\qquad Only defines $(\varpi,\mu)\to\gamma_G$ not inverse\\  
\\ 
Decoupled \cite{Bean:2009wj}&\qquad$\eta=\frac{1}{1+\varpi}$
&\qquad$\gamma_G=\frac{\ln(\dot\Delta_m/\mathcal{H}\Delta_m)}{\ln\Omega_m(a)}$
&\qquad Over-specified (also enforces Poisson eqn).\\
\end{tabular}
\caption{
Translation between several different parameterizations of modified gravity 
and the $\varpi\mu$CDM framework.  
}
\label{comparisontable}
\end{table*}

Since none of these model-independent approaches start from an action, 
one must be careful to trace the system of equations to make sure that 
the phenomenological modifications do not under- or over-specify the 
system\,\footnote{
A careful reading of \cite{Bean:2009wj} reveals that there four unknowns -- 
$\phi$, $\psi$, $\delta$, and $\theta$ -- are evolved with five equations 
-- the continuity equation (\ref{SE1}), Euler equation (\ref{SE2}), Poisson 
equation (\ref{poisson}), and the post-GR parameter equations 
$\phi=\eta\psi$ and $\dot\Delta_m=\mathcal{H}\Delta_m\Omega_m^{\gamma_G}$.
Thus, the system is over-specified.
}
and do satisfy stress-energy conservation. 
Another approach involves testing consistency relations 
valid in GR between observables; see for example 
\cite{Zhang:2005gh,Zhang:2007nk, Song:2008xd}.

\section{Influence of Gravity Modifications on Observations \label{sec:influ}} 

The behavior of the CMB, weak lensing, and matter power spectrum in the 
$\varpi$CDM scenario have been discussed in 
\cite{Daniel:2008et, Daniel:2009kr}. The consequences are slightly different 
when we introduce $\mu$ in the $\varpi\mu$CDM parameterization. In the case 
that $\varpi<0$ and $\mu<1$, both lead to an amplification of low-$\ell$ CMB 
power; $\varpi>0$ and $\mu>1$ both suppress it.  This allows us to play the 
two parameters against each other, combining positive (negative) values of 
$\varpi$ with smaller (larger) values of $\mu$ to generate non-GR power 
spectra that appear 
to be in better agreement with the data than those obtained within the 
confines of the $\varpi$CDM model. That either parameter can enhance or 
suppress power results in a degeneracy between $\varpi$ and $\mu$ in any 
multi-parameter exploration of the data.  Observations that can break this 
degeneracy therefore become vital to diagnosing departures from GR.  

For the purposes of the discussion in this section, we will assume the 
redshift dependences 
\begin{eqnarray}
\varpi&=&\varpi_0 a^3\nonumber\\
\mu&=&1+\mu_0a^3\label{paramsacubed} \,.
\end{eqnarray}
Note this form can be motivated by the scaling argument in 
\cite{Linder:2007hg}, that the deviations in the expansion history should 
keep pace with the deviations in the growth history.  Otherwise one tends 
to either violate GR at early times (causing difficulties for primordial 
nucleosynthesis and the CMB) or does not achieve acceleration by the present. 
In addition to the CMB, we also discuss the effects of our post-GR parameters 
on the matter power spectrum and on weak lensing statistics.

\subsection{CMB Anisotropy Spectrum \label{sec:cmb}}

\begin{figure}[!t]
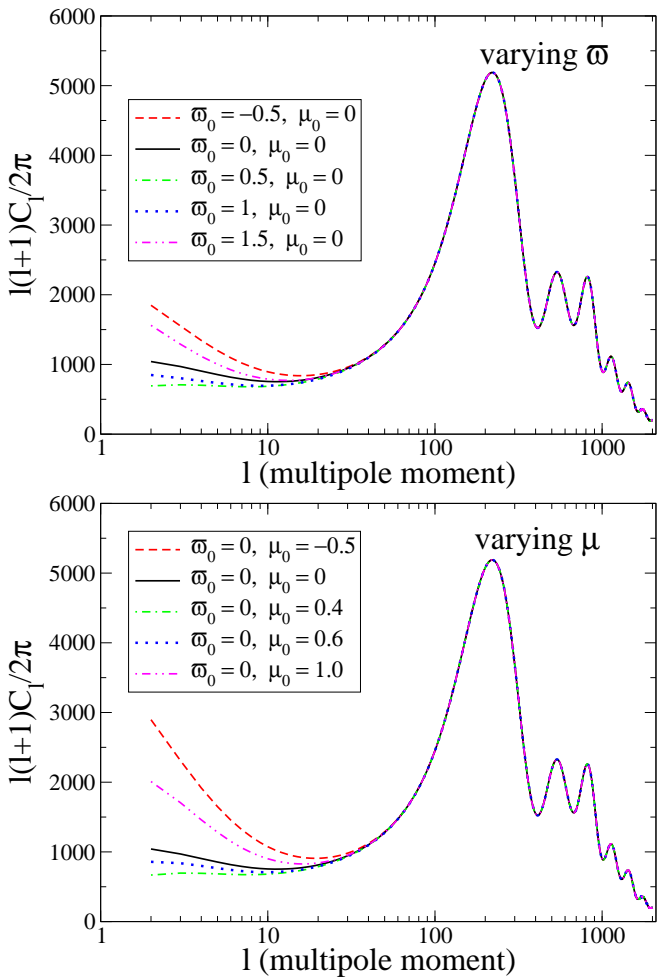

\includegraphics[width=\columnwidth]{fig_1a.eps}
\label{clvarpiacubed}
\includegraphics[width=\columnwidth]{fig_1b.eps}
\label{clmuacubed}
\caption{
CMB anisotropy spectra are plotted as a function of the parameters 
$\varpi_0$ and $\mu_0$ in Eqs.~(\ref{paramsacubed}).  
As in 
\cite{Daniel:2008et}, the post-GR effects all occur in the low-$\ell$ 
multipoles.  The CMB anisotropy is more sensitive to variations in $\mu_0$ 
than to variations in $\varpi_0$.   See Fig.~\ref{quadrupoleplot} for more 
on this point and on varying $\varpi_0$  and $\mu_0$ simultaneously. 
}
\label{clacubedfig}%
\end{figure}

We modified versions of the public Boltzmann codes CMBfast \cite{Seljak:1996is} and CAMB \cite{Lewis:1999bs} to evolve the cosmological perturbations according to parameterization (\ref{paramsacubed}) and the equations of motion
presented in Sec.~\ref{sec:vmucdm}.  We used these codes to generate examples 
of CMB anisotropy and matter power spectra for different values of $\varpi_0$ 
and $\mu_0$; in order to focus on the non-GR effects, in this section 
all other cosmological 
parameters are set to their WMAP5 maximum likelihood values \cite{wmapparams}.
Figure~\ref{clacubedfig} shows the
resulting CMB anisotropy spectra.  As in \cite{Daniel:2008et} for $\varpi$CDM, 
negative values and extreme positive values of the post-GR parameters amplify 
the power in the low-$\ell$ multipoles.  Moderate positive values suppress the 
low-$\ell$ power.  This is a manifestation of the integrated Sachs-Wolfe (ISW) 
effect.  The high-$\ell$ power is unaffected. 

The ISW effect arises when time evolving $\phi$ and $\psi$ potentials cause 
a net energy shift in CMB photons.  The CMB ISW power is sourced as 
\beq 
C_l \sim(\dot\phi+\dot\psi)^2 \,. \label{eq:iswterm}
\eeq 
As was discussed in \cite{Daniel:2009kr}, the evolution of $\phi$ and $\psi$ 
potentials in the universe is a competition between gravitational collapse 
trying to deepen the potentials and cosmic expansion trying to dilute them.   
Under GR with a cosmological constant, the expansion wins and the source term 
for the ISW $\dot\phi+\dot\psi>0$ (note $\phi$, $\psi<0$). 
By weakening gravity, $\varpi_0$ or $\mu_0<0$ tilts the competition even  more 
towards cosmic expansion, hastening the dilution of $\phi$ and $\psi$, 
causing $\dot\phi+\dot\psi$ to be even larger, and amplifying the ISW effect.  
Positive $\varpi_0$ or  $\mu_0$ amplifies gravity -- either by directly 
deepening the Newtonian potential $\psi$ so that mass is more attractive 
($\varpi_0>0$ case) 
or by  causing $\Delta_m$ to source a deeper potential through the modified 
Poisson equation ($\mu_0>0$ case) -- so that the dilution due to cosmic 
expansion is slowed, leading to a weaker ISW effect.  In the case of 
extremely positive $\varpi_0$ or $\mu_0$ the ISW  deepening is so pronounced 
that the sign of $\dot\phi+\dot\psi$ is reversed, but since the  ISW effect 
in the power spectrum depends on the square, the ISW effect is again 
amplified.  High-$\ell$ power is unaffected because the ISW is a sub-dominant 
effect on those scales. 

\begin{figure}[!t]
\center
\includegraphics[width=\columnwidth]{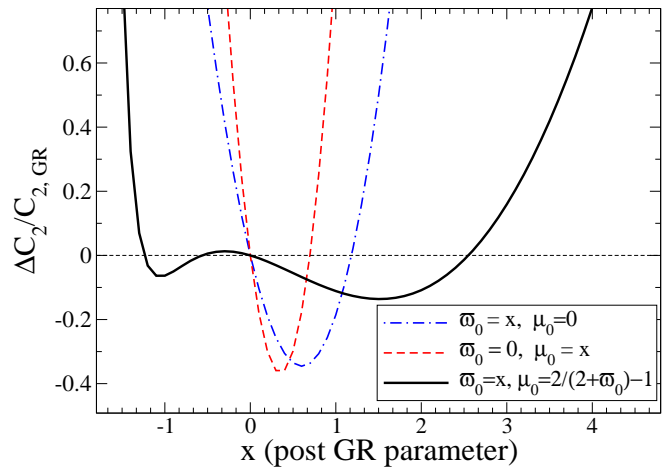}
\caption{
The change in quadrupole power relative to the value in GR is plotted as 
a function of $\varpi_0$ and $\mu_0$.  The blue, dot-dashed curve shows 
the effects of varying $\varpi_0$ with fixed $\mu_0=0$.  The red, dashed 
curve shows the effects of varying $\mu_0$ with fixed $\varpi_0=0$.  One 
can mimic the unmodified GR CMB spectrum
over a much wider range of post-GR parameter values 
by simultaneously varying $\varpi_0$ 
and $\mu_0$ in opposite directions, as shown in the black, solid curve 
using $\mu_0=2/(2+\varpi_0)-1$.  The horizontal dotted line 
denotes perfect agreement with GR. 
}
\label{quadrupoleplot}
\end{figure}

Figure~\ref{quadrupoleplot} more clearly illustrates this bimodal behavior 
by plotting the change in quadrupole power relative to GR as a function of 
the post-GR parameter, varying one at a time  (compare Fig.~4 in 
\cite{Daniel:2009kr}).  The blue, dot-dashed curve is  generated by varying 
$\varpi_0$ and holding fixed $\mu_0=0$.   The red, dashed curve is generated 
by varying $\mu_0$ and holding fixed $\varpi_0=0$. Note that the CMB appears 
to be more sensitive to differing values of $\mu_0$ than of $\varpi_0$.  The 
black, solid curve is generated by varying $\varpi_0$ and compensating for 
this by setting $\mu_0=2/(2+\varpi_0)-1$.  This choice is motivated by the 
alternative definition of the unmodified Poisson equation 
\begin{equation}
\label{altpoissoneqn}
-k^2(\phi+\psi)/2=4\pi G a^2\bar{\rho}_m\Delta_m 
\end{equation} 
(see further discussion in the next section).  
We see that, for a wide range of values of $\varpi_0$,
complementary ($\varpi_0>0$ and $\mu_0<0$ or vice-versa) values of $\mu_0$
cancel out much of the late-time ISW effect found in 
Fig.~\ref{clacubedfig}, as alluded to in the introduction to this section.

\subsection{Matter Power Spectrum and Weak Lensing Statistics \label{sec:mat}}

\begin{figure}[!h]
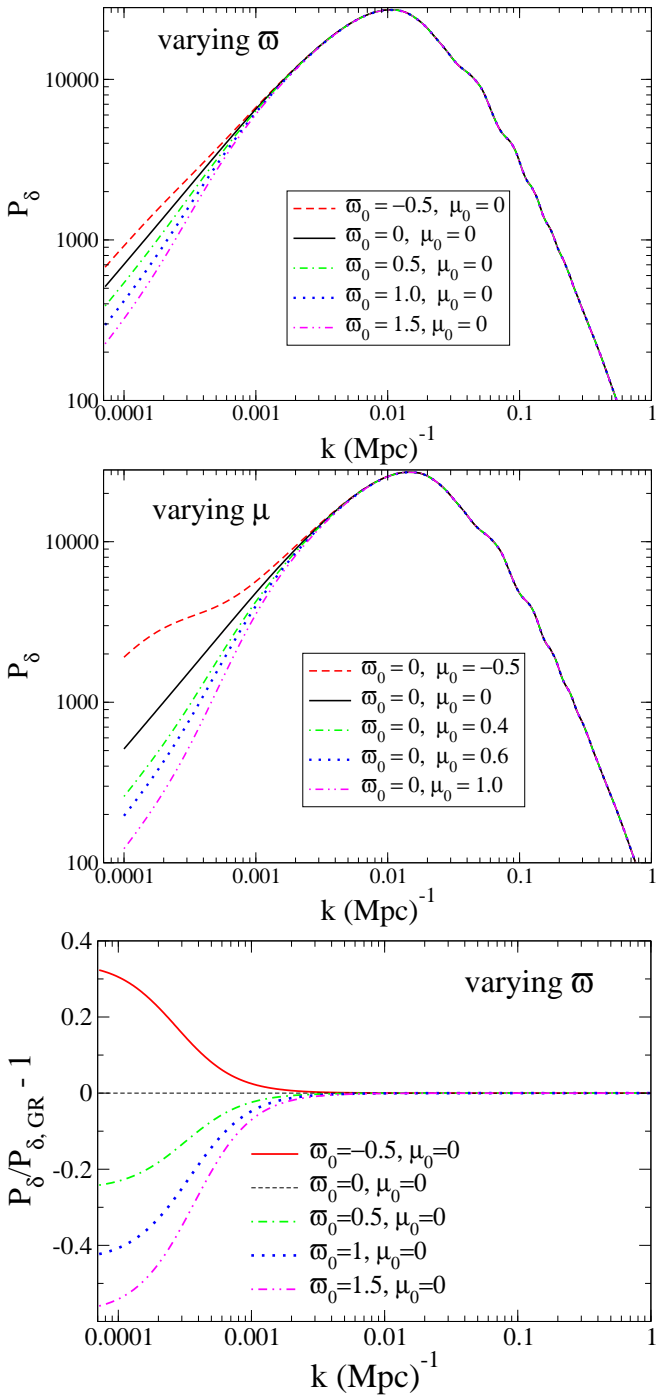

\includegraphics[width=\columnwidth]{fig_3a.eps}
\label{pspcvarpiacubed}
\includegraphics[width=\columnwidth]{fig_3b.eps}
\label{pspcmuacubed}
\includegraphics[width=\columnwidth]{fig_3c.eps}
\caption{
We plot the matter power spectrum (normalized to $k=1\,{\rm Mpc}^{-1}$) 
generated by varying the parameters 
$\varpi_0$ and $\mu_0$.  Unlike under $\varpi$CDM \cite{Daniel:2008et}, 
even our scale-independent parameterization has scale-dependent effects 
due to the $k^2$ factor in the Poisson equation. The bottom panel shows 
the residuals of the top panel, i.e.\ the deviation relative to GR when 
varying $\varpi_0$ (the $\mu_0$ case looks similar), to highlight the 
scale-dependent regime at low $k$ and scale-independent regime at high $k$. 
}
\label{pspcacubedfig} 
\end{figure}

We investigate the power spectrum of the matter perturbations $\delta_m$ as a 
function of wavenumber $k$ in Fig.~\ref{pspcacubedfig} for the same set of 
models.  Again the most dramatic post-GR effects occur at large scales.  This 
is not due to any scale dependence in the modifications (we took $\varpi$ 
and $\mu$ to be independent of $k$), but simply from the $k^2$ factor in the 
modified Poisson equation (\ref{mudef}). 

For the weak lensing shear correlation function, as for many other 
observables, we need to know how overdensities grow with scale factor.  In 
the case of GR and $\varpi$CDM, this is a relatively simple proposition 
since the growth of overdensities $\Delta_m$ is scale-independent after 
decoupling.  As just discussed, this no longer holds for $\varpi\mu$CDM.
It is possible, using energy conservation and Eqs.~(\ref{varpidef}) and 
(\ref{mudef}), to derive a second-order differential equation for the 
evolution of $\Delta_m$. 
We show the derivation and result in the Appendix, and focus here on the 
parameter dependence. 

With the exception of one term on the middle line of Eq.~(\ref{Ddd}), all 
of the terms containing metric potential modifications to general relativity 
(the $\mu$ and $\varpi$ terms) are multiplied by a factor of 
$\mathcal{H}^2/k^2$.  Hence, we expect that the strongest departures from GR 
predictions occur for small values of $k$.  
Since the most important aspect for comparing modifications against 
observations is the change in shape of the power spectrum, rather 
than its normalization, in Figure~\ref{pspcacubedfig} we normalize the 
power spectrum to agree with $P_{GR}$ at large $k$. 
The strongest deviation in shape indeed occurs for 
$k\lesssim0.002\,{\rm Mpc}^{-1}$. 
The one exceptional term in Eq.~(\ref{Ddd}) 
is precisely the $(1+\varpi)\mu$ term discussed 
in Sec.~\ref{sec:gamma} entering the gravitational growth index $\gamma_G$ 
formalism, and this will dominate for large values of $k$, giving a 
scale-independent enhancement (suppression) for positive (negative) 
$\varpi_0$ or $\mu_0$.

\begin{figure}[!t]
\center
\includegraphics[width=\columnwidth]{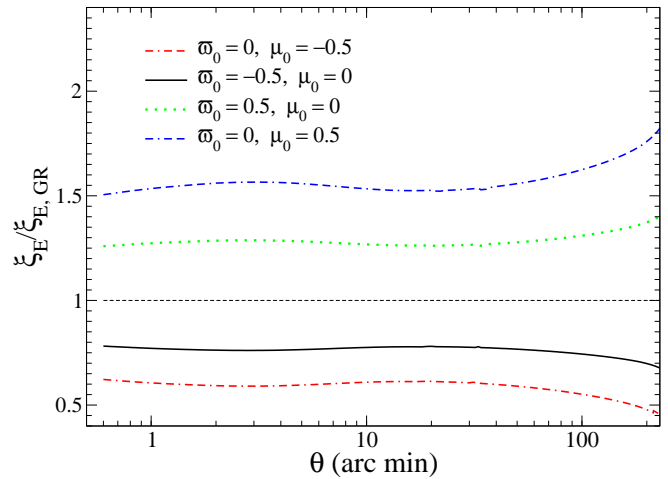}
\caption{
We plot the ratio of the E mode of the weak lensing shear two-point correlation function (Eq.~8 of \cite{Fu:2007qq}) to the same statistic calculated in GR, with all parameters but either $\varpi_0$ or $\mu_0$ fixed, to see the influence of the non-GR parameters.  For the most part, post-GR parameters serve to renormalize the correlation function.  As with the CMB anisotropy and matter power spectra, the effect is more sensitive to changes in $\mu_0$ than to changes in $\varpi_0$.
}
\label{wklnfig}
\end{figure}

Figure~\ref{wklnfig} plots $\xi_E$, the E mode of the weak lensing shear 
two-point correlation function (Eq.~(8) of Ref.~\cite{Fu:2007qq}), normalized 
to the value under GR as a function of angular separation on the sky.  
For the 
angular scales of interest, the effects of changing $\mu_0$ and $\varpi_0$ 
principally manifest themselves as a renormalization of $\xi_E$.  This is 
because the scales plotted are much smaller than the scales 
($k\sim \mathcal{H}$) at which shape-changing effects manifested themselves in 
Figure \ref{pspcacubedfig}. 
Non-linear power is treated using the usual subroutine halofit based on the
semi-analytic fitting scheme presented in Ref.~\cite{Smith:2002dz}.
While we acknowledge that this is not strictly appropriate for modified gravity,
we have no reason to think that the effect will be substantial for reasonable
values of $\mu_0$ and $\varpi_0$.  Furthermore, the constraints
presented below in Section \ref{sec:obs} appear to principally 
derive from effects at the low-$k$,
rather than the high-$k$, limit.

\section{Constraints on Deviations from GR \label{sec:obs}} 

We now examine constraints imposed by current data on deviations from GR, 
allowing a large set of cosmological parameters to vary simultaneously.  
The investigation includes two different functional dependences for the 
gravitational modification parameters $\varpi(a)$ and $\mu(a)$. 
The first model for the post-GR 
parameter form does not assume a particular redshift dependence but 
allows $\varpi$ and $\mu$ to take independent values in each of three 
redshift bins.  (In fact, we slightly smooth the transitions so as to 
avoid infinities in the derivatives entering the ISW effect, with a 
transition modeled by an arctan form of width $\Delta a=0.01$.)  
That is,  $\mu=\{1+\mu_{0a},1+\mu_{0b},1+\mu_{0c}\}$ and 
$\varpi=\{\varpi_{0a},\varpi_{0b},\varpi_{0c}\}$ for $\{2< z\leq 9, 
1<z\leq 2,z\leq 1\}$.  We assume that $\varpi$ and $\mu$ are 
scale-independent.  For $z>9$ we assume that differences from GR are 
negligible so $\mu=1$ and $\varpi=0$. 

We test this theory against the data using a modified version of the public 
MCMC code COSMOMC \cite{Lewis:1999bs,Lewis:2002ah,COSMOMC_notes} with a 
module (first presented in \cite{Lesgourgues:2007te}) to incorporate the 
COSMOS weak lensing tomography data \cite{Massey:2007gh} and data from the 
CFHTLS survey \cite{Fu:2007qq}.  We also include WMAP5 CMB data 
\cite{Dunkley:2008ie,Nolta:2008ih,Hinshaw:2008kr} 
and Union2 supernova distance data \cite{amanullah}. 
In all cases, we use the full covariance matrix (including systematics 
in the Union2 case) provided by the group who
collected and initially analyzed the data.
In addition to the post-GR parameters, the parameter set includes 
$\Omega_bh^2$, $\Omega_ch^2$, $\theta$ (the ratio of the sound horizon
to the angular diameter distance to last scattering), $\tau$ (the optical
depth to reionization), $n_s$, the amplitude of the SZ effect,
and the amplitude of primordial scalar perturbations.  We assume that
$w=-1$ for our effective dark energy, that $\Omega_K=0$, and that there
are no massive neutrinos contributing to dark matter.
Each weak lensing data set requires 3 nuisance parameters.
Thus, we integrate over a total of up to 16 parameters, depending
on the data sets used and the parametrization of $\mu$ and $\varpi$ chosen.
Under the binned parametrization, we vary $\mu$ or $\varpi$ but not both simultaneously, which would require 19 independent parameters.
This choice was made both for convenience and to reproduce the analysis
of Ref.~\cite{Bean:2009wj}.  Additional MCMC calculations done in which
both $\mu$ and $\varpi$ were allowed to vary in all three bins also returned
results consistent with GR in the presence of a cosmological constant.

\begin{figure}[!t]
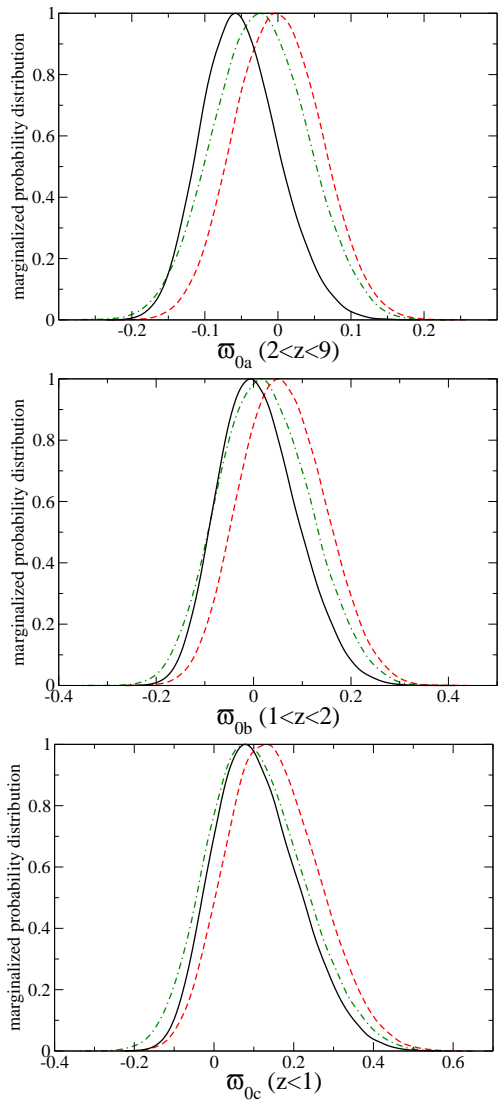

\includegraphics[scale=0.27]{fig_5a.eps}
\label{varpi0a_nomu}
\includegraphics[scale=0.27]{fig_5b.eps}
\label{varpi0b_nomu}
\includegraphics[scale=0.27]{fig_5c.eps}
\label{varpi0c_nomu}
\caption{
Marginalized probabilities of the post-GR parameters $\varpi_{0a,b,c}$ 
defined in high, medium and low redshift bins respectively.
The parameter $\mu$ has been fixed to $\mu=1$, consistent with 
General Relativity.  Green (dot-dashed) 
curves are constraints determined from the WMAP 5 year \cite{Dunkley:2008ie} 
and supernova Union2 \cite{amanullah} data sets only. Red (dashed) curves 
also include the COSMOS weak lensing tomography data \cite{Massey:2007gh}. 
Black (solid) curves use measurements of the aperture mass taken from 
the CFHTLS weak lensing survey \cite{Fu:2007qq} in addition to COSMOS, 
WMAP5, and Union2.
}
\label{nomu1Dfig} 
\end{figure}

\begin{figure}[!t]
\center
\includegraphics[width=\columnwidth]{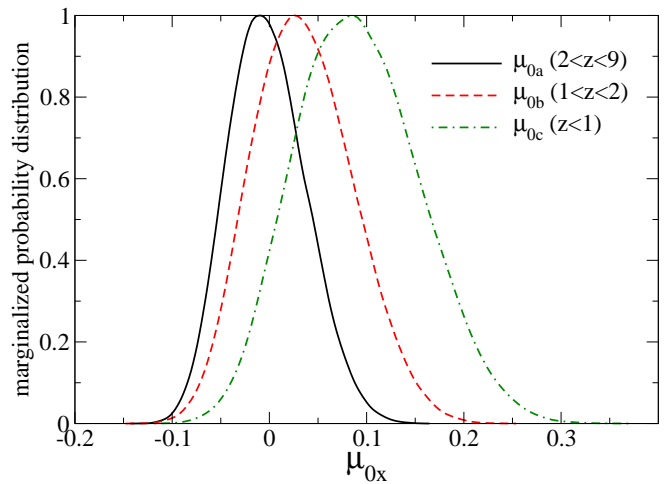}
\caption{
Marginalized probabilities of the post-GR parameters $\mu_{0a,b,c}$ 
defined in high, medium, and low redshift bins respectively.  
The parameter $\varpi$ has been fixed to $\varpi=0$, consistent with
General Relativity.
All curves show constraints derived using data from
the WMAP 5 year release \cite{Dunkley:2008ie}, 
supernova Union2 set \cite{amanullah}, and COSMOS \cite{Massey:2007gh} plus
CFHTLS \cite{Fu:2007qq} weak lensing data. 
}
\label{mu1Dfig}
\end{figure}

Figure~\ref{nomu1Dfig} shows the marginalized probabilities on the 
$\varpi_{0a,b,c}$ parameters for runs in which $\mu_{0a,b,c}=0$, so that 
$\mu=1$ and the Poisson equation defined as in Eq.~(\ref{poisson}) remains 
valid at all redshifts.
Figure \ref{mu1Dfig} shows similar constraints on $\mu_{0a,b,c}$ 
in the case where $\varpi_{0a,b,c}=0$ .  
Our results in all cases are consistent with GR 
within the 95\% confidence limit, although they do allow the possibility 
of departures from GR with $\varpi_0$ or $\mu_0\sim0.1$.

Constraints on the usual cosmological parameters are largely unaffected by 
the introduction of $\mu$ and $\varpi$.  Mean values shift by less than 
$\sim1\sigma$ and marginalized uncertainties are comparable between GR and
non-GR MCMC runs.  The only notable exceptions are 
$\sigma_8$ and $\Omega_c h^2$ (the physical density of cold dark matter 
in the universe), whose marginalized uncertainties increase by up to 
a factor 2.3 upon the introduction of post-GR 
parameters.  This is consistent with the observation that $\mu$ and $\varpi$
principally modify the growth history of cosmological perturbations.

\begin{figure}[!h]
\center
\includegraphics[width=\columnwidth]{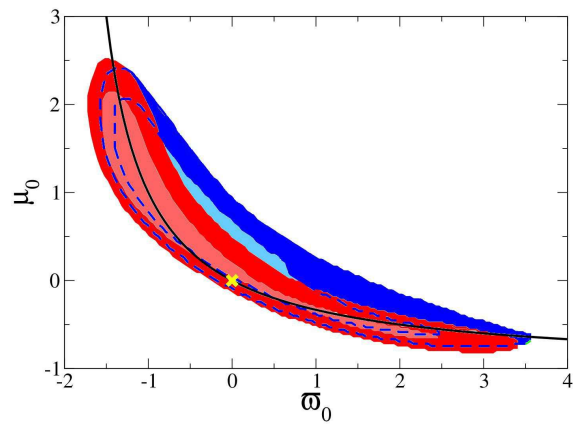}
\caption{68\% (inner) and 95\% (outer) confidence contours for modified 
gravity parameters are plotted in the $\varpi_0$--$\mu_0$ plane 
(all other parameters marginalized),   
where $\varpi_0$ and $\mu_0$ are defined as in Eq.~(\ref{paramsacubed}). 
Blue (background) contours use the WMAP 5 year, supernova 
Union2, and COSMOS data sets.  Red (foreground) contours use WMAP 5 year, 
supernova Union2, COSMOS and CFHTLS data.  The black curve plots the 
degeneracy direction $\mu_0=2/(2+\varpi_0)-1$ from Figure~\ref{quadrupoleplot}.  The yellow x denotes GR parameter values.
}
\label{contourplot}
\end{figure}

Figure~\ref{contourplot} plots the 2-dimensional confidence contours for the 
post-GR parameters $\varpi_0$, $\mu_0$ in the case of redshift dependence
as in Eqs.~(\ref{paramsacubed}).  Note that since this parameterization 
has the strongest effect at low redshift, the greater sky area of CFHTLS 
has more leverage in constraining the parameters than the greater depth 
of COSMOS.
For the binned parametrization, the constraints from MCMC runs with
WMAP5+Union2+CFHTLS (no COSMOS) were indistinguishable from those including
COSMOS as well, supporting the supposition that the sky coverage of CFHTLS is,
for current data, more important than the redshift depth of COSMOS in
constraining the post-GR parameters.

Table~\ref{constrainttable} presents the 95\% constraints on our post-GR
parameters for all of the MCMC calculations considered in Figures
\ref{nomu1Dfig}-\ref{contourplot}.  All of the results are consistent with GR.

\begin{table*}[!t]
\begin{tabular}{l l l l l}
Binned $\varpi$, $\mu=1$:&&\qquad Binned $\mu$, $\varpi=0$:&\qquad 
Parameterization (\ref{paramsacubed}):&\\
COSMOS&\qquad+CFHTLS&\qquad+CFHTLS&\qquad COSMOS&\qquad+CFHTLS\\
\hline
$-0.11<\varpi_{0a}<0.12$&\qquad$-0.15<\varpi_{0a}<0.060$&\qquad$-0.074<\mu_{0a}<0.080$&
\qquad$-1.4<\varpi_0<2.8$&\qquad$-1.6<\varpi_0<2.7$\\
$-0.098<\varpi_{0b}<0.23$&\qquad$-0.13<\varpi_{0b}<0.18$&\qquad$-0.058<\mu_{0b}<0.14$&
\qquad$-0.67<\mu_0<2.0$&\qquad$-0.83<\mu_0<2.1$\\
$-0.054<\varpi_{0c}<0.39$&\qquad$-0.074<\varpi_{0c}<0.33$&\qquad$-0.023<\mu_{0c}<0.22$
\end{tabular}
\caption{
95\% confidence limits on post-GR parameters in the MCMC calculations
considered in Figures~\ref{nomu1Dfig} (left two columns), \ref{mu1Dfig} 
(middle column), and \ref{contourplot} (right two columns).  Columns labeled 
``COSMOS'' use WMAP5+Union2+COSMOS data.  Columns labeled ``+CFHTLS'' use
WMAP5+Union2+COSMOS+CFHTLS data.  Recall that in the binned parameterization,
redshift bin $a$ is $2<z<9$, redshift bin $b$ is $1<z<2$, and redshift bin 
$c$ is $z<1$.
}
\label{constrainttable}
\end{table*}

We also note that in Figure~\ref{contourplot} the contours exhibit the same 
degeneracy implied by Figure~\ref{quadrupoleplot}.  Apparently, the probe 
of growth provided by current weak lensing data is not able to add much more 
leverage to the CMB data. 
This can also be seen in the lack of significant change in the width of 
the probability distributions in Figure~\ref{nomu1Dfig} when adding 
weak lensing. 

The degeneracy illustrated in Figure~\ref{quadrupoleplot} is plotted as the 
black, solid curve in Figure~\ref{contourplot}.  The agreement with the 
likelihood contours is quite interesting, 
calling to mind the discussion in Sec.~\ref{sec:cmb} about parameter 
covariances.  This arose from the observation that an unmodified Poisson 
equation (\ref{altpoissoneqn}) that relates the {\it sum} of the two 
metric potentials to the 
underlying density fluctuations leaves the large-scale CMB predictions 
nearly unchanged when varying the ratio of the metric potentials, i.e.\ 
$\varpi$.  That 
degeneracy is due to the fact that the large-scale CMB predictions depend 
on the sum of the two metric potentials (cf.~Eq.~\ref{eq:iswterm}).  If this 
sum is directly related to the underlying density perturbation then the 
only effect $\varpi$ can have on the large-scale CMB is through its effect 
on the evolution of $\Delta_m$; by contrast, if the Poisson equation is 
of the form of Eq.~(\ref{mudef}), where only one potential is related to 
$\Delta_m$, then $\varpi$ also appears in a multiplicative factor.  Thus 
the specific approaches to modifying gravity give distinct relations between 
the parameters and the observables. 

For observations that depend on the combination $\phi+\psi$ there will be 
a degeneracy along the curve (see Eqs.~\ref{eq:poisson_ppf} and 
\ref{translation1}) 
\begin{equation}
\mu = \frac{2}{2+\varpi}\,. \label{eq:compensate} 
\end{equation}
We find numerically that this degeneracy applies approximately to both 
large-scale CMB as well as weak lensing observations, even though both 
measurements have a further dependence on $\varpi$ and $\mu$ through the 
growth factor (cf.~Eq.~\ref{eq:gammawu}).  The relation in 
Eq.~(\ref{eq:compensate}) gives the black, solid curve in 
Figure~\ref{contourplot} and indeed is quite close to a degeneracy in 
the constraints.


\section{Discussion \label{sec:fut}}

Testing general relativity on cosmological length scales is an 
exciting prospect enabled by improvements in data.  To interpret such 
a test requires an approach to parameterizing modifications from GR, 
similar to the PPN method for tests within the solar system and using 
binary pulsars, but appropriate for cosmic scales.  Numerous 
parameterizations have been suggested and we compare, and in some 
cases, unify them through a ``translation'' table.  
These approaches can effectively be interpreted within one formalism with two 
parameters $\varpi$ and $\mu$ (an extension of the previous $\varpi$CDM 
scenario). 

In this generalized $\varpi\mu$CDM model, even if the two parameters 
characterizing modifications to gravity are scale independent we find 
effects that are visible in the large-scale structure matter power spectrum, 
and thus in weak lensing shear correlations, that depend on scale.  
We give quantitative results for the effects of the modifications on the 
cosmic microwave background temperature power spectrum, the growth of
matter density perturbations and the density power spectrum, and the weak 
lensing statistics, along with analysis of the physical basis of the effects. 
On large scales in the density power spectrum, values of $\varpi$ or 
$\mu$ above their GR values cause suppression of power
while leading to enhancement on smaller scales. 

We confront the modifications to GR with current cosmological observations, 
analyzing CMB (WMAP 5-year), supernovae (Union2), and weak lensing (CFHTLS 
and COSMOS) data.  Employing two different forms of dependence of the 
modifications on redshift, we find no evidence at 95\% confidence level for 
such extensions to GR, regardless of the combinations of data used.  Note 
that this holds for both the data employed by \cite{Bean:2009wj} (which 
used an overspecified system of equations in that analysis), and a more 
comprehensive set of observations. 

We also verify the trade-off between $\varpi$ and $\mu$ predicted 
analytically.  Such covariance leads to an interesting degeneracy 
for measurements depending on the sum of the metric potentials, although 
growth measurements depend on a different combination. Since large
scale CMB and weak lensing depend on the sum of the
metric potentials, one could consider the Poisson equation
for the sum, and here the key parameter is the effective
NewtonÕs constant $\tilde{G}_{\rm eff}=\mu(2+\varpi)/2$. 
The matter density growth factor is primarily sensitive to extensions 
beyond GR in terms of the factor $\Sigma=\mu(1+\varpi)$. 
These parameters still appear to have covariance, however, in 
our initial explorations. Overall, this suggests 
that exploration of gravity
through cosmological measurements requires a sufficiently
flexible theory space and a diverse set of observations.

As seen from Figures~\ref{quadrupoleplot}-\ref{wklnfig}, robust 
identification of deviations from GR will require measurement over a 
large range of scales.  Well below the Hubble scale, the modifications 
we have examined become scale independent and so can become confused 
with shifts in the fiducial amplitude ($\sigma_8$), galaxy bias, or 
normalization errors from photometric redshift estimation of weak lensing 
source densities.  These will need to be addressed to have confidence in 
claims of any detected deviation, as will allowance for expansion histories 
different from $\Lambda$CDM. 

Finally, future data, including observations sensitive to growth and the 
growth rate, and those sensitive to the expansion history, will be essential 
to providing true tests of the framework of gravity on cosmic scales.

\acknowledgments

We are extremely grateful to the Supernova Cosmology Project for permission 
to use the Union2 supernova data before publication.  
We acknowledge the sue of the Legacy Archive for Microwave Background 
Data Analysis (LAMBDA).  Support for LAMBDA is provided by the NASA Office of
Space Science.
We thank Rachel Bean for helpful discussions on Ref.~\cite{Bean:2009wj}. 
This work has been supported by the World Class University grant
R32-2008-000-10130-0 (SD, EL). 
EL has been supported in part by the
Director, Office of Science, Office of High Energy Physics, of the
U.S.\ Department of Energy under Contract No.\ DE-AC02-05CH11231.
AC acknowledges support from NSF CAREER AST-0645427. 
RC acknowledges support from NSF CAREER AST-0349213.
LL was supported by the Swiss National Science Foundation under 
Contract No. 2000 124835 1.

\appendix
\section{Density Perturbation Growth} 

We here obtain the analog of the GR second order differential equation for 
matter density perturbation evolution, working in the conformal Newtonian 
gauge.  After matter-radiation decoupling, conservation of energy gives 
\begin{eqnarray}
\dot\delta_m&=&3\dot\phi-\theta_m\label{deltadot}\\
\dot\theta_m&=&-\mathcal{H}\theta_m+k^2\psi \,,\label{thetadot}
\end{eqnarray} 
assuming $\delta p=\sigma=0$, i.e.\ there is no pressure, no pressure 
perturbation, and no anisotropic shear.  Rearranging Eq.~(\ref{deltadot}) 
and substituting Eq.~(\ref{thetadot}), we can write 
\begin{eqnarray}
\theta_m&=&3\dot\phi-\dot\delta_m\nonumber\\
&=&3\dot\phi-\dot\Delta_m
+\frac{d}{d\tau}\left(\frac{3\mathcal{H}\theta_m}{k^2}\right)
\nonumber\\
&=&3\dot\phi-\dot\Delta_m
+\frac{3}{k^2}\left[\theta_m\dot{\mathcal{H}}-\mathcal{H}^2\theta_m+\mathcal{H}k^2\psi\right]
\nonumber\\
&=&\frac{3\dot\phi-\dot\Delta_m+3\mathcal{H}\psi}{1-3(\dot{\mathcal{H}}-\mathcal{H}^2)/k^2} \,. 
\label{thetadef} 
\end{eqnarray}

We can use Eqs.~(\ref{mudef}) and (\ref{varpidef}) to write $\psi$ in
terms of $\Delta_m$, $\mu$, $\varpi$, and background quantities; similarly 
we can use the time derivative of Eq.~(\ref{mudef}) to write $\dot\phi$. 
This gives 
\begin{eqnarray}
\dot\Delta_m&=&\dot\delta_m+\frac{3\dot{\mathcal{H}}\theta_m}{k^2}+
\frac{3\mathcal{H}\dot\theta_m}{k^2}\nonumber\\
&=&3\dot\phi-\theta_m+\frac{3\dot{\mathcal{H}}\theta_m}{k^2}
+\frac{3\mathcal{H}}{k^2}
\left(-\mathcal{H}\theta_m+k^2\psi\right) \,, \label{Ddotdef}
\end{eqnarray}
where the second equality comes from using Eqs.~(\ref{deltadot}) 
and (\ref{thetadot}).  Substituting Eq.~(\ref{thetadef}) into 
(\ref{Ddotdef}) would just return the truism $\dot\Delta_m=\dot\Delta_m$.
However, if we take the first conformal time derivative of 
Eq.~(\ref{Ddotdef}) before substituting, we find a second order 
differential equation describing the evolution of $\Delta_m$ for arbitrary
$\varpi(a,k)$ and $\mu(a,k)$.  We omit the explicit 
copious algebra and show the result: 

\begin{widetext}
\begin{eqnarray}
\ddot\Delta_m\left(1+\frac{3}{k^2}\Gamma\mu\right)
&=&\dot\Delta_m\Bigg\{-\frac{3}{k^2}
\left(2\dot\Gamma\mu+2\Gamma\dot\mu\right)
-\frac{3}{k^2}(1+\varpi)\Gamma\mu\mathcal{H}
+\left(\mathcal{H}+3\frac{\ddot{\mathcal{H}}}{k^2}-9\frac{\mathcal{H}\dot{\mathcal{H}}}{k^2}+3\frac{\mathcal{H}^3}{k^2}\right)
\times\frac{-3\Gamma\mu-k^2}{k^2-3\left(\dot{\mathcal{H}}-\mathcal{H}^2\right)}
\Bigg\}\nonumber\\
&&+\Delta_m\Bigg\{-\frac{3}{k^2}\left(\ddot\Gamma\mu+\Gamma\ddot\mu+2\dot\Gamma\dot\mu\right)
-(1+\varpi)\frac{\Gamma\mu}{k^2}\left(-k^2+6\dot{\mathcal{H}}-3\mathcal{H}^2\right)
-3\dot\varpi\frac{\Gamma\mu}{k^2}\mathcal{H}
-\frac{3}{k^2}(1+\varpi)\mathcal{H}\left(\dot\Gamma\mu+\Gamma\dot\mu\right)
\nonumber\\
&&\phantom{\Delta_m\Bigg\{}
+\left(\mathcal{H}+3\frac{\ddot{\mathcal{H}}}{k^2}-9\frac{\mathcal{H}\dot{\mathcal{H}}}{k^2}+3\frac{\mathcal{H}^3}{k^2}\right)
\times\frac{-3\left(\dot\Gamma\mu+\Gamma\dot\mu\right)-3(1+\varpi)\mathcal{H}\Gamma\mu}{k^2-3\left(\dot{\mathcal{H}}-\mathcal{H}^2\right)}
\Bigg\} \,,
\label{Ddd}
\end{eqnarray}
\end{widetext}
where $\Gamma=4\pi Ga^2\bar\rho_m$. 
All that we have assumed in this derivation is that matter and $\Lambda$ 
are the only constituents of the background cosmology so that 
Eqs.~(\ref{deltadot}) and (\ref{thetadot}) hold.



\begin{thebibliography}{99}

\bibitem{Frieman:2008sn}
  J.~Frieman, M.~Turner and D.~Huterer,
  Ann.\ Rev.\ Astron.\ Astrophys.\  {\bf 46}, 385 (2008)
  [arXiv:0803.0982 [astro-ph]].

\bibitem{Caldwell:2009ix}
  R.~R.~Caldwell and M.~Kamionkowski,
  Ann.\ Rev.\ Nucl.\ Part.\ Sci.\  {\bf 59}, 397 (2009)
  [arXiv:0903.0866 [astro-ph.CO]].

\bibitem{Massey:2007gh}
R.~Massey {\it et al.},
Astrophys.\ J.\ Suppl.\ {\bf 172}, 239 (2007)
[arXiv:astro-ph/0701480].

\bibitem{Fu:2007qq}
L.~Fu {\it et al.},
Astron.\ Astrophys.\ {\bf 479} 9 (2008)
[arXiv:0712.0884 [astro-ph]].

\bibitem{Bertschinger:2006aw}
  E.~Bertschinger,
  Astrophys.\ J.\  {\bf 648}, 797 (2006)
  [arXiv:astro-ph/0604485].

\bibitem{Caldwell:2007cw}
  R.~Caldwell, A.~Cooray and A.~Melchiorri,
  Phys.\ Rev.\  D {\bf 76}, 023507 (2007)
  [arXiv:astro-ph/0703375].

\bibitem{Zhang:2007nk}
  P.~Zhang, M.~Liguori, R.~Bean and S.~Dodelson,
  Phys.\ Rev.\ Lett.\  {\bf 99}, 141302 (2007)
  [arXiv:0704.1932 [astro-ph]].

\bibitem{Amendola:2007rr}
  L.~Amendola, M.~Kunz and D.~Sapone,
  JCAP {\bf 0804}, 013 (2008)
  [arXiv:0704.2421 [astro-ph]].

\bibitem{Hu:2007pj}
  W.~Hu and I.~Sawicki,
  Phys.\ Rev.\  D {\bf 76}, 104043 (2007)
  [arXiv:0708.1190 [astro-ph]].

\bibitem{Amin:2007wi}
  M.~A.~Amin, R.~V.~Wagoner and R.~D.~Blandford,
  arXiv:0708.1793 [astro-ph].

\bibitem{Jain:2007yk}
  B.~Jain and P.~Zhang,
  Phys.\ Rev.\  D {\bf 78}, 063503 (2008)
  [arXiv:0709.2375 [astro-ph]].

\bibitem{Bertschinger:2008zb}
  E.~Bertschinger and P.~Zukin,
  Phys.\ Rev.\  D {\bf 78}, 024015 (2008)
  [arXiv:0801.2431 [astro-ph]].


\bibitem{Hu:2008zd}
  W.~Hu,
  Phys.\ Rev.\  D {\bf 77}, 103524 (2008)
  [arXiv:0801.2433 [astro-ph]].


\bibitem{Song:2008vm}
  Y.~S.~Song and K.~Koyama,
  JCAP {\bf 0901}, 048 (2009)
  [arXiv:0802.3897 [astro-ph]].

\bibitem{Schmidt:2008hc}
  F.~Schmidt,
  Phys.\ Rev.\  D {\bf 78}, 043002 (2008)
  [arXiv:0805.4812 [astro-ph]].

\bibitem{Skordis:2008vt}
  C.~Skordis,
  Phys.\ Rev.\  D {\bf 79}, 123527 (2009)
  [arXiv:0806.1238 [astro-ph]].

\bibitem{Linder:2009kq}
  E.~V.~Linder,
  Phys.\ Rev.\  D {\bf 79}, 063519 (2009)
  [arXiv:0901.0918 [astro-ph.CO]].

\bibitem{Dent:2009wi}
  J.~B.~Dent, S.~Dutta and L.~Perivolaropoulos,
  Phys.\ Rev.\  D {\bf 80}, 023514 (2009)
  [arXiv:0903.5296 [astro-ph.CO]].

\bibitem{Schmidt:2007vj}
  F.~Schmidt, M.~Liguori and S.~Dodelson,
  Phys.\ Rev.\  D {\bf 76}, 083518 (2007)
  [arXiv:0706.1775 [astro-ph]].

\bibitem{Zhao:2008bn}
  G.~B.~Zhao, L.~Pogosian, A.~Silvestri and J.~Zylberberg,
  Phys.\ Rev.\  D {\bf 79}, 083513 (2009)
  [arXiv:0809.3791 [astro-ph]].

\bibitem{Pogosian:2010tj}
  L.~Pogosian, A.~Silvestri, K.~Koyama and G.~B.~Zhao,
  arXiv:1002.2382 [astro-ph.CO].

\bibitem{Song:2008xd}
  Y.~S.~Song and O.~Dore,
  JCAP {\bf 0903}, 025 (2009)
  [arXiv:0812.0002 [astro-ph]].

\bibitem{Serra:2009kp}
  P.~Serra, A.~Cooray, S.~F.~Daniel, R.~Caldwell and A.~Melchiorri,
  Phys.\ Rev.\  D {\bf 79}, 101301 (2009)
  [arXiv:0901.0917 [astro-ph.CO]].

\bibitem{Zhao:2009fn}
  G.~B.~Zhao, L.~Pogosian, A.~Silvestri and J.~Zylberberg,
  Phys.\ Rev.\ Lett.\  {\bf 103}, 241301 (2009)
  [arXiv:0905.1326 [astro-ph.CO]].

\bibitem{Guzik:2009cm}
  J.~Guzik, B.~Jain and M.~Takada,
  Phys.\ Rev.\ D {\bf 81}, 023503 (2010) 
  [arXiv:0906.2221 [astro-ph.CO]].

\bibitem{Kosowsky:2009nc}
  A.~Kosowsky and S.~Bhattacharya,
  Phys.\ Rev.\  D {\bf 80}, 062003 (2009)
  [arXiv:0907.4202 [astro-ph.CO]].

\bibitem{Beynon:2009yd}
  E.~Beynon, D.~J.~Bacon and K.~Koyama,
  Mon.\ Not.\ Roy.\ Astron.\ Soc.\ {\bf 403}, 353 (2010)
  arXiv:0910.1480 [astro-ph.CO].

\bibitem{Masui:2009cj}
  K.~W.~Masui, F.~Schmidt, U.~L.~Pen and P.~McDonald,
  Phys.\ Rev.\ D {\bf 81}, 062001 (2010)
  arXiv:0911.3552 [astro-ph.CO].

\bibitem{Song:2010rm}
  Y.~S.~Song, L.~Hollenstein, G.~Caldera-Cabral and K.~Koyama,
  arXiv:1001.0969 [astro-ph.CO].

\bibitem{Zhang10} 
W.~Cui, P.~Zhang and X.~Yang, 
arXiv:1001.5184 [astro-ph.CO] 




\bibitem{Di Porto:2007ym}
  C.~Di Porto and L.~Amendola,
  Phys.\ Rev.\  D {\bf 77}, 083508 (2008)
  [arXiv:0707.2686 [astro-ph]].

\bibitem{Nesseris:2007pa}
  S.~Nesseris and L.~Perivolaropoulos,
  Phys.\ Rev.\  D {\bf 77}, 023504 (2008)
  [arXiv:0710.1092 [astro-ph]].

\bibitem{Dore:2007jh}
  O.~Dore {\it et al.},
  arXiv:0712.1599 [astro-ph].

\bibitem{Daniel:2008et}
  S.~F.~Daniel, R.~R.~Caldwell, A.~Cooray and A.~Melchiorri,
  Phys.\ Rev.\  D {\bf 77}, 103513 (2008)
  [arXiv:0802.1068 [astro-ph]].

\bibitem{Smith:2002dz}
R.~E.~Smith {\it et al.} [The Virgo Consortium Collaboration],
Mon.\ Not.\ Roy.\ Astron.\ Soc.\ {\bf 341}, 1311 (2003)
[arXiv:astro-ph/0207664].

\bibitem{Yamamoto:2008gr}
  K.~Yamamoto, T.~Sato and G.~Huetsi,
  Prog.\ Theor.\ Phys.\  {\bf 120}, 609 (2008)
  [arXiv:0805.4789 [astro-ph]].

\bibitem{Daniel:2009kr}
  S.~F.~Daniel., R.~R.~Caldwell, A.~Cooray, P.~Serra, A.~Melchiorri, 
  Phys.\ Rev.\  D {\bf 80}, 023532 (2009)
  [arXiv:0901.0919 [astro-ph.CO]].

\bibitem{Giannantonio:2009gi}
  T.~Giannantonio, M.~Martinelli, A.~Silvestri and A.~Melchiorri,
  arXiv:0909.2045 [astro-ph.CO].
 
\bibitem{Bean:2009wj}
  R.~Bean,
  arXiv:0909.3853 [astro-ph.CO].
 
\bibitem{Bertotti:2003rm}
  B.~Bertotti, L.~Iess and P.~Tortora,
  Nature {\bf 425}, 374 (2003).

\bibitem{Shapiro:2004zz}
  S.~S.~Shapiro, J.~L.~Davis, D.~E.~Lebach and J.~S.~Gregory,
  Phys.\ Rev.\ Lett.\  {\bf 92}, 121101 (2004).

\bibitem{Taylor:1994zz}
  J.~H.~Taylor,
  Rev.\ Mod.\ Phys.\  {\bf 66}, 711 (1994).

\bibitem{Lyne:2004cj}
  A.~G.~Lyne {\it et al.},
  Science {\bf 303}, 1153 (2004)
  [arXiv:astro-ph/0401086].

\bibitem{Will:1993ns}
  C.~M.~Will,
  ``Theory and experiment in gravitational physics,''
{\it  Cambridge, UK: Univ. Pr. (1993) 380 p}

\bibitem{Will:2005va}
  C.~M.~Will,
  Living Rev.\ Rel.\  {\bf 9}, 3 (2006)
  [arXiv:gr-qc/0510072].

\bibitem{Ma:1995ey}
  C.~P.~Ma and E.~Bertschinger,
  Astrophys.\ J.\  {\bf 455}, 7 (1995)
  [arXiv:astro-ph/9506072].

\bibitem{Turner:2001mx}
  M.~S.~Turner and A.~G.~Riess,
  Astrophys.\ J.\  {\bf 569}, 18 (2002)
  [arXiv:astro-ph/0106051].

\bibitem{Linder:2005in}
  E.~V.~Linder,
  Phys.\ Rev.\  D {\bf 72}, 043529 (2005)
  [arXiv:astro-ph/0507263].

\bibitem{Linder:2007hg}
  E.~V.~Linder and R.~N.~Cahn,
  Astropart.\ Phys.\  {\bf 28}, 481 (2007)
  [arXiv:astro-ph/0701317].

\bibitem{Zhang:2005gh}
  P.~Zhang,
  Astrophys.\ J.\  {\bf 647}, 55 (2006)
  [arXiv:astro-ph/0512422].

\bibitem{Seljak:1996is}
  U.~Seljak and M.~Zaldarriaga,
  Astrophys.\ J.\  {\bf 469}, 437 (1996)
  [arXiv:astro-ph/9603033].

\bibitem{Lewis:1999bs}
  A.~Lewis, A.~Challinor and A.~Lasenby,
  Astrophys.\ J.\  {\bf 538}, 473 (2000)
  [arXiv:astro-ph/9911177].

\bibitem{wmapparams}
 \verb|http://lambda.gsfc.nasa.gov/product|\\ 
 \verb|/map/dr3/params/lcdm_sz_lens_wmap5.cfm|

\bibitem{Lewis:2002ah}
  A.~Lewis and S.~Bridle,
  Phys.\ Rev.\  D {\bf 66}, 103511 (2002)
  [arXiv:astro-ph/0205436].

\bibitem{COSMOMC_notes}
A.~Lewis and S.~Bridle, \url{http://cosmologist.info/notes/COSMOMC.ps.gz}

\bibitem{Lesgourgues:2007te}
J.~Lesgourgues, M.~Viel, M.~G.~Haehnelt and R.~Massey,
JCAP {\bf 0711}, 008 (2007)
[arXiv:0705.0533 [astro-ph]].

\bibitem{Dunkley:2008ie}
  J.~Dunkley {\it et al.}  [WMAP Collaboration],
  Astrophys.\ J.\ Suppl.\ {\bf 180}, 306 (2009)
  arXiv:0803.0586 [astro-ph].

\bibitem{Nolta:2008ih}
M.~R.~Nolta {\it et al.} [WMAP Collaboration],
Astrophys.\ J.\ Suppl.\ {\bf 180}, 296 (2009)
[arXiv:0803.0593 [astro-ph]].

\bibitem{Hinshaw:2008kr}
G.~Hinshaw {\it et al.} [WMAP Collaboration],
Astrophys.\ J.\ Suppl.\ {\bf 180}, 225 (2009)
[arXiv:0803.0732 [astro-ph]].

\bibitem{amanullah} 
R.~Amanullah {\it et al.},
Astrophys.\ J.\ {\bf 716}, 712 (2010)
[arXiv:1004.1711 [astro-ph.CO]].

\end{thebibliography}
\end{document}